\documentclass[12pt,reqno]{article}
\usepackage{amsmath,amssymb,amsfonts,mathtools} 
\usepackage{amssymb}  	 
\usepackage{makeidx}  	 
\usepackage{graphicx} 		 
\usepackage{epsfig}
\usepackage{appendix}		
\usepackage{empheq}
\usepackage{setspace} 	
\usepackage{enumitem}
\usepackage[sort&compress,numbers]{natbib}
\usepackage{pstricks}
\usepackage{soul}  
\usepackage[dvips]{curves}  
\usepackage{xcolor}
\usepackage{listings}
\usepackage{amsbsy}

\usepackage{multicol}
\usepackage{verse}
\usepackage{pstricks, pst-text}                  
\usepackage{color}
\usepackage{fancyhdr}
\usepackage[T1]{fontenc}						
\usepackage{newpxtext,newpxmath}   	
\usepackage{titlesec}
\titleformat{\section}{\normalfont\fontsize{12}{13}\bfseries}{\thesection}{6pt}{\setstretch{0.1}}
  \titleformat{\subsection}
  {\normalfont\fontsize{11}{12}\bfseries}{\thesubsection}{1em}{}
  \titleformat{\subsubsection} 
  {\normalfont\fontsize{11}{11}\bfseries}{\thesubsubsection}{1em}{}

\newcommand\myfigure[5]{%
  \ifdim#2>.8\linewidth
    {%
      \centering
      \includegraphics[width=#3]{#4}%
      \captionof{figure}{#5}%
    }%
  \else
  \begin{wrapfigure}{#1}{#2}
    \includegraphics[width=#3]{#4}
    \caption{#5}
  \end{wrapfigure}
  \fi
}

\usepackage{graphicx}
\usepackage{wrapfig}
\usepackage{caption}
\newcommand{\linkcolor}{blue}
\usepackage[colorlinks=true,linkcolor=\linkcolor,urlcolor=\linkcolor]{hyperref} 
\usepackage{academicons}
\definecolor{orcidlogocol}{HTML}{A6CE39}

\def\d{{\,\rm d}}
\def\e{{\rm e}}

\begin{document}
\noindent
\textbf{\Large 100+ years of colossal confusion on colloidal coagulation. Part I: Smoluchowski's work on absorbing boundaries}\\[2em]
K. Razi Naqvi (\url{razi.naqvi@ntunu.no})\\
\textit{Department of Physics, Norwegian University of Science and Technology (NTNU), 7094 Trondheim, Norway}\\
\vspace*{2ex}

\leftskip=20truept
\rightskip=20truept
\setstretch{0.9}
{\small
\noindent
\textbf{\textit{Abstract}}---A report by Brillouin (from Perrin's laboratory) on the rate of adsorption of `granules' to a glass plate [\textit{Ann. Chim. Phys.} 27 (1912) 412--23] prompted Marian von Smoluchowski (MvS) to interpret the data in terms of his newly developed theory of restricted Brownian motion. Placing an adsorbing wall at $x=0$, he modelled the particle concentration $n(x,t)$ as that solution of the diffusion equation which vanished at the wall, a boundary condition (BC) hereafter called SBC. A gaping discrepancy between his theory and Brillouin's data elicited a suggestion from MvS (that a particle might not adhere to the wall on every impact), but no further action---other than that of applying his theory to spherically symmtric systems. In a paper written before, but published shortly after MvS's untimely death [\textit{Proc. Roy. Acad. Amst.} 20 (1918) 642--58], H. C. Burger erected a new and sturdier framework, which led him to an alternative BC, $D(\partial n/\partial x)_{x=0}=\varkappa n(0,t)$, applicable to a surface with an arbitrary absorption probability ($1\leq\varepsilon\leq 0$); a fallacy (that subsequently claimed more victims, including the present author) prevented him from deducing the correct expression for $\varkappa$. Burger's approach became ``The Road Not Taken'', while the SBC became the cornerstone of colloidal coagulation and bimoleculer reaction kinetics. Burger's approach (but not the ABC) was partly rediscovered by Kolmogorov, and used by Sveshnikov and Fuchs.The emended version of Burger's BC is shown here to coincide with that deduced from the Klein-Kramers equation [\textit{Phys. Rev. Lett.} \textbf{49} (1982) 304--07; \textit{J. Chem. Phys.} \textbf{78} (1983) 2710--12] and the Lorentz model of random flights [\textit{J.  Phys. Chem.} \textbf{86} (1982) 4750--56].} \\[1em]


\setstretch{1.04}
\leftskip=0truept
\rightskip=0truept
\vspace{2em}

\begin{multicols}{2}
\poemtitle{\small Burger warns against SBC}
\settowidth{\versewidth}{There was an old party of Lyme}
\begin{verse}[\versewidth]
Herman Burger once did firmly say,\\
"Never, NEVER use the SBC, I pray"; \\
\vin Yet each \textit{JCP} and \textit{JPC} we see, \\
\vin Refers still to Smoluchowski.\\
\end{verse}

\section{Introduction}
\label{sec:Intro}
Presumably Marian von Smoluchowski (d. September 1917) never saw a paper (read April 1917; published 1918) whose author, H. C. Burger, a young associate of L. S. Ornstein, began his paper (which had no abstract) with the following introductory statement \cite{Burger1918PRAcadAmst}:\par
\smallskip
\leftskip=20truept
\setstretch{0.9}
{\small\noindent
The purpose of this inquiry is of two kinds viz:\par
\noindent 1$^{\rm st}$. to give a rigorous proof of some well known formulae of
the Brownian movement, with and without external force.\par
\noindent 2$^{\rm nd}$. to trace the boundary-condition at a fixed wall and to interpret some experimental results concerning this.
}\par
\setstretch{1.04}
\leftskip=0truept
\smallskip 
Much to the detriment of scholars occupied with the theoretical discourse of colloidal coagulation and other sundry diffusion-mediated reactions, this thoughtful contribution by Burger (hereafter HCB1918) failed to captivate the attention of those who would have benefited most from it, even though he gave it what must have been an eye-catching title at the time, namely ``On the Theory of the Brownian Movement and the experiments of Brillouin''. 

Chandrasekhar, who acquird willy-nilly---as a reward for his 1943 review \cite{Chandrasekhar1943RMP}---the mantle of the chief expositor (in English) of Smoluchowski's ideas on Brownian motion and colloidal coagulation, included HCB1918 in the ``Bibliographical Notes'', but never explicitly referred to it in the text; he might have cited the article without reading even the first five lines, or without assenting to Burger's arguments, especially the contention (in which $F$ denotes the concentration of the diffusing species): ``Strictly speaking the bondary-condition can \ldots never be $F(0,t) = 0$ [as proposed by Smoluchowski], because then $\varkappa$ must be infinite, which is impossible.'' Burger's comment was based on his conclusion that, for a black wall, $\varkappa=DF^\prime(0,t)/F(0,t)=\overline{v}/4$,  where $\overline{v}=(8k_{\bf B}T/\pi m)^{1/2}$ is the average thermal speed of the diffusing particle.  

In 1976, Sunagawa and Doi \cite{SunagawaDoi1975} drew attention to HCB1918 in a sigle sentence: ``Physically, this is equivalent to including the fractional reflection of the reactive sites upon collision and can be expressed by the radiation boundary condition''; of the two references cited by them at this point, one is HCB1918, the other a 1949 article \cite{Collins1949Kimball}, authored by Collins and Kimball (C\&K). An alert, well-informed and history-sensitive reader would not fail to notice that Burger had recommended the radiation boundary condition (RBC) some thirty years before C\&K (who too failed to link the constant $\varkappa$ to meaningful physical quantities).

The third and the last work which refers to HCB1918 came to my notice after the completion of the present article. A search for a more recent reference to HCB1918, or to one that I had failed to notice during previous rounds, dredged up Chen-Pang Yeang's \textit{Transforming Noise} \cite[pp. 224-28]{Yeang2023TransformNoise}, in which Burger receives due attention and a just applause, but only for accomplishing the first (and, in my view, less demanding) of his avowed aims. Yeang covers \S\S1--2 of HCB1918, which focus on the derivation of the diffusion equation (DE) in a field-free system as well as one where a field is present, but it says nothing about  \S\S3--4, where Burger develops a new framework for the kinetics of diffusion-mediated reactions and wrestles with the the derivation of the boundary conditions.
 
The problem of inferring the correct boundary condition (BC) to be imposed on a solution of the DE at a partially or totally absorbing surface has been handled from two different and mutually incompatible viewpoints. One approach, pioneered by Marian von Smoluchowski (MvS for short), sets the concentration of the diffusing species equal to zero at a perfectly absorbing surface \cite{Smoluchowski1916PhysZeit}; the introduction of the RBC, purportedly applicable to any surface (perfectly or partly absorbing, or non-absorbing), is usually credited to C\&K \cite{Collins1949Kimball}. This line of reasoning, which has dominated the literature on chemical kinetics, will henceforth be called the SCK approach. 

An altogether different strategy is to start with a particle transport equation (PTE), together with an appropriate boundary condition, for describing the temporal and spatial variation of the distribution function in phase space, and proceed to extract, in a consistent manner, a configuration space description in terms of the DE and the associated boundary condition; the accuracy of the results furnished by the DE---the crudest approximation in a PTE analysis---can be assessed by going to a higher-order approximation.\par
 At present, it would be enough for us to note that the results obtained by treating the DE as an approximation (to some PTE) disagree with those obtained by using the SCK approach. This is an alarming situation, all the more so because the community of chemical kineticists seem to have turned a deaf ear to the alarm bells, which have been ringing intermittently for over a century. Alarming, but not unique, because wrinkles are easily formed in the fabric of science, as in any other fabric, but they are not as easily ironed out as in most real fabrics. In science, dewrinkling entails convincing members of a community that the arguments which are dear to them are logically indefensible. The aversion to abandoning a long-held view was well diagnosed by Oliver Heaviside in the preface to his whimsical and engaging \textit{Electromagnetic Theory} \cite[pp. x--xi]{Heaviside1893ElectromagTheory1}:\par
 \medskip
\leftskip=20truept
\setstretch{0.85}
{\small\noindent
It is not long since it was taken for granted that the common electrical units were correct. That curious and obtrusive constant $4\pi$ was considered by some to be a sort of blessed dispensation, without which all electrical theory would fall to pieces. I believe \ldots that the $4\pi$  was an unfortunate and mischievous mistake, the source of many evils. In plain English, the common system of electrical units involves an irrationality of the same kind as would be brought into the metric system of weights and measures, were we to define the unit area to be the area, not of a square with unit side, but of a circle of unit diameter. The constant $\pi$ would then obtrude itself into the area of a rectangle, and everywhere it should not be, and be a source of great confusion and inconvenience. So it is in the common electrical units, which are truly irrational. Now, to make a mistake is easy and natural to man. But that is not enough. The next thing is to correct it. When a mistake has once been started, it is not necessary to go on repeating it for ever and ever with cumulative inconvenience.}\par
\medskip
\leftskip=0truept
\setstretch{1.04}
                                     
A step towards correcting some chronic errors was taken in 1979, when the present author published a short note \cite{KRN1979CPLwidespread}, which bore the title
``Concerning some widespread errors in diffusion-controlled reaction kinetics'', and conveyed a simple message: The literature on the subject is replete with relations which imply that $a=b$ and $b=c$, but $a\neq c$.  The point was made by displaying four equations (numbered 1, 8--10), all taken from the \textit{same} paper, and adding:
``It follows from the first three equations that \ldots, which cannot be reconciled with eq. (10).''  The only prerequisites for grasping the message were familiarity with elementary algebra and the willingness to read the short note. Short or long, right or wrong (actually the former), this note made no impact and has so far received only two citations. \par 

In the early 1980s, serveral groups \cite{Harris1981JCP,Harris1983Sphere,Burschka+Titulaer1981JSPa,Titulaer1984JStatPhys,Kneller+Titulaer1984Physica,Mayya+Sahni1983JCP,Chaturvedi+Agarwal1983ZfP}, including the present author and two of his Trondheim colleagues (the Trondheim group, for short), began applying the Klein-Kramers and other transport equations to Milne's problem (and some of its variants), the name given to the prototypical problem in the present context; the Trondheim group, especially interested in establishing and extending the limits of applicability of the DE, brought this part of the project to a successful end \cite{KRN+KJM+SW1982JCP,KRN+KJM+SW1982PRL,KRN1982CPL-RW1,KRN1982CPL-RW2,KRN+SW+KJM1982JPC,KRN+SW+KJM1983JCP,SW+KJM+KRN1983PhysRevA,KRN1984Colloid}. After a corpus of consistent results had been obtained \cite{Marshall+Watson1985Drop,Menon+Sahni1985PhysRevA,Menon+Kumar+Sahni1986Physica,Kneller+Titulaer1985Physica,WidderTitulaer1989JStatPhysV55,WidderTitulaer1989JStatPhysV56,Kainz+Titulaer1992TwoStream}, the activity tapered off under the decelerating influence of the law of diminishing returns. Unfortunately, most kineticists seem to be unaware of what was achieved in that period and what remains to be done during the next spurt (if any). In view of the importance of the topic, it seems worthwhile to present a short tutorial review, making as few mathematical demands on the reader as the topic would permit. Though provision of some new insight, rather than of new findings, is the principal aim of this article, the reader is assured that what is presented below amounts to more than a mere paraphrase of existing results and arguments.

\section{Note on terminology and mathematical symbols}

The phrase \textit{Brownian motion} will not be treated here as a mere synonym for random displacements of all kinds. When speaking of \textit{regular} Brownian motion and \textit{inverse} Brownian motion, I will adhere to the definitions formulated by some previous authors 
\cite{Keilson+Storer1952,Andersen+ShulerJCP1964,Hoare1970Nature,Hoare1971AdvChemPhys}. The ratio, $\gamma_{\raisebox{-2pt}{\hspace{-2pt}\scriptsize $m$}}\equiv m_{u}/m_{v}$, between the masses, $m_{u}$ and $m_{v}$, of the sol$u$te (or particle) and sol$v$ent molecules (or host), lies at the root of the difference between the two kinds of particle transport.  For our purpose, Brownian motion and its inverse (which will be called \textit{Lorentzian flight}) may be identified with the limits $\gamma_{\raisebox{-2pt}{\hspace{-2pt}\scriptsize $m$}}\rightarrow\infty$ and $\gamma_{\raisebox{-2pt}{\hspace{-2pt}\scriptsize $m$}}\rightarrow 0$, respectively. The importance of these extremes resides in the fact that the corresponding kinetic equations become linear and, in each limit, exact results are available for the Milne problem, which serve as yardsticks for results pertaining to other systems.

The PTEs for the two limits $\gamma_{\raisebox{-2pt}{\hspace{-2pt}\scriptsize $m$}}\rightarrow\infty$ and $\gamma_{\raisebox{-2pt}{\hspace{-2pt}\scriptsize $m$}}\rightarrow 0$ are known as the Lorentz-Boltzmann (or the linear Boltzmann) equation (LBE) and the Klein-Kramers equation (KKE), respectively.  Alternative appellations are, for the former, the equation of radiative transfer (or one-velocity Boltzmann equation for neutron transport) and, for the latter, the {\it generalized} Fokker-Planck equation, but the italicized qualifier is often dropped.  A third PTE \cite{BGK1954PhysRev}, named the BGK kinetic equation, has also proved to be exceedingly useful in elucidating the issues treated in this article.

\subsection{Einstein's derivation of the DE (1905)}
\label{subsec:EinsteinDE}

In his 1905 article \cite{Einstein1905AnP}, Einstein uses (twice) the phrase ``solute molecules (or suspended particles)'', and once ``solute molecules, or suspended bodies (hereinafter termed briefly `particles')''. In \S3, entitled ``Theory of the Diffusion of Small Spheres in Suspension'', Einstein uses Stokes's law for the drag force $F=6\pi\eta av$ on a sphere (of radius $a$) moving with a velocity $v$ in a suspending medium of viscosity $\eta$, and Fick's law $D=-\partial n/\partial x$ (without naming either eponym), arrives at his Eq. (7),
\vspace{-4pt}
\begin{equation}\label{eq:StokesEinst}
D=\frac{{\textsf{\textit R}}T }{N_{\rm A}}\frac{1}{6\pi\eta\,a}, \tag{E-7}
\vspace{-4pt}
\end{equation}
and ends \S2 with the remark: ``The coefficient of diffusion of the suspended substance [a Brownian particle] therefore depends (except for universal constants and the absolute temperature) only on the coefficient of viscosity of the liquid and on the size of the suspended particles.'' The symbols {\textsf{\textit{R}}} and $N_{\rm A}$ stand for the gas constant and the Avogadro constant, respectively. Here we will express Eq. (\ref{eq:StokesEinst}) in the more general form
\vspace{-4pt}
\begin{equation}\label{eq:EinstZeta}
D=\frac{k_{\bf B}T}{m\zeta},\quad(\mbox{$\zeta$: friction coefficient}).
\vspace{-4pt}
\end{equation}
Einstein's derivation of the DE was based on a model that did not discriminate between different particle-host systems  \cite[\S4]{Einstein1905AnP,Einstein1956Investigations}.  He related the particle concentration $n(x, t+\tau)$  to the product $n(x+s,t)\,\phi(s;\tau)$, where $\phi(s;\tau)$ is the probability that a particle will undergo a displacement $s$ during an intrval of duration $\tau$:
\vspace{-4pt}
\begin{equation}
n(x,t+\tau)=\int_{-\infty}^\infty\hspace{-2pt} \d s\, n(x+s,t)\phi(s;\tau)
\vspace{-4pt}
\end{equation}
The limits of integration make it patent that he was thinking of an \textit{unrestricted}, one-dimensional random walk. After specifying the statistical properties of $\phi(s;\tau)$, imposing a restriction on $\tau$, and dropping $\partial^k n/\partial t^k$ ($k\geq 2$) and $\partial^k m/\partial x^m$ ($m\geq 3$), he arrived at the DE, with $D=\overline{s^2}/2\tau$, $\overline{s^2}$ being the mean squared displacement in time $\tau$.  That the fundamental solution as well as $\phi$ turned out to be gaussians is an illustration of the central limit theorem, because the interval $\tau$ is assumed to be so large that every individual displacement $s$ is a resultant of a very large number of random displacements executed by the particle under observation.  

\subsection{Smoluchowski's notion of the mean free path of a Brownian particle (1906)}
\label{subsec:SmoluDE}
MvS, being fully conversant with the kinetic theory of gases, found it natural to forge his theory of Brownian motion by using the statistical approach cultivated by Maxwell and Boltzmann, and by making full use of Jeans's idea of `persistence of velocities after collision' \cite[pp. 236--41]{Jeans1904Dynamical}. Since a detailed discussion of two of his 1906 papers \cite{Smoluchowski1906Cracovie,Smoluchowski1906Annalen} falls outside the scope of the present article, I recall some words from Sommerfeld's obituary of Smoluchowski \cite{Sommerfeld1917PhysZeit}: \par
 \medskip
\leftskip=20truept
\setstretch{0.85}
{\small\noindent
The Smoluchowski and Einstein derivations each have their own particular advantages. Smoluchowski lets us look deeper into the mechanism of the collisions; the dependence on the temperature, the particle size and the viscosity of the suspending agent is reliably derived; only the numerical coefficient remains uncertain due to the difficulty of averaging. Einstein boldly takes control of the final result without having to dwell on the details of the process; doubtlessly, Einstein's expression, which differs from Smoluchowski's by 27/64, is correct.}\par
\medskip
\leftskip=0truept
\setstretch{1.04}

The opening remarks of the last section of the second 1906 paper \cite{Smoluchowski1906Annalen} are of particular relevance to us:\par
\medskip
\leftskip=20truept
\setstretch{0.85}
{\small\noindent
The result of the previous section can be summarized as follows: particles suspended in a liquid or gaseous medium behave as if they were independent gas molecules with normal kinetic energy, but with an exceedingly small mean free path \ldots }\par
\medskip
\leftskip=0truept
\setstretch{1.04}
The task of expressing the diffusion coefficient of a particle in terms of the apparent free path ($\ell$) of a Brownian particle in a manner that is completely consistent with the generalized Einstein relation, Eq. (\ref{eq:EinstZeta}), was accomplished in 1982 and led to the relation \cite{KRN+KJM+SW1982JCP,KRN+KJM+SW1982PRL}: $\ell=(9\pi/8)^{1/2}\Lambda$, with $\Lambda=D/(k_{\bf B}T/m)^{1/2}$.

\subsection{Essential mathematical ingredients}
\label{subsec:ingredients}
The DE, $\partial n(x,t)=D(\partial^2 n/\partial x^2$, can be deduced, as the reader well knows, by combining the exact relation,
\vspace{-4pt}
\begin{equation}\label{eq:EoCont}
\frac{\partial n}{\partial t} + \frac{\partial j}{\partial x} =0,
\vspace{-4pt}
\end{equation}
known as the equation of continuity, with the empirical (and approximate) relation,
\vspace{-4pt}
\begin{equation}\label{eq:Fick01}
j=-D \frac{\partial n}{\partial x}, 
\vspace{-4pt}
\end{equation}
often called Fick's law, between the particle current density $j$ and the concentration gradient. In the works where it is named Fick's first law, the DE is called Fick's second law.

Finally we introduce the symbols $j{\hspace{-2pt}\smash[t]{\strut}}_{\pm}(x,t)$ for the two partial fluxes in the $\pm x$-direction across an elementary reference surface plane or spherical), and note that
\begin{equation}\label{eq:jfullsumjparts}
j(x,t)  = j{\hspace{-2pt}\smash[t]{\strut}}_{+}(x,t) -  j{\hspace{-2pt}\smash[t]{\strut}}_{-}(x,t).
\end{equation}
The terms \textit{flux} and \textit{current} will be treated as synonyms.
\section{Brillouin's expriment on colloids}
\label{sec:BrillouinExpt}
In his landmark book \textit{Les Atomes} \cite{Perrin1913Atomes}, the English translation of which, entitled \textit{The Atoms}, appeared in 1916 \cite{Perrin1916Atoms}, Perrin described Brillouin's measurements in such detail (pp. 184--88; pp. 129--33) as to make it unnecessary to look at Brillouin's paper \cite{Brillouin1912AnnChimPhys}.
Perrin had observed that, under some conditions, a large colloidal particle became immobile after striking the wall of the cover glass, which resulted in a decrease in the concentration of the particles in the vicinity of the absorbing wall. Working in his laboratory, Brillouin used this property to determine the diffusion coefficient ($D$) of the particles, which was supposed to satisfy Eq. (\ref{eq:StokesEinst}). Brillouin counted $M$, the number of trapped particles at different times $t$, and fitted his data to the relation 
\vspace{-4pt}
\begin{equation}
M(t) = {{\textstyle\frac{1}{2}}}n_0\Delta_{\rm r} =
{{\textstyle\frac{1}{2}}}\,n_0\,\sqrt{2Dt}=n_0\,\sqrt{\frac{Dt}{2}}, \label{eq:Brillo1}
\vspace{-4pt}
\end{equation}
in which $n_0\equiv n(x,0)$ denotes the initial number density of the particles (the product of the concentration and the cross sectional area of the sample holder); he deduced Eq. (\ref{eq:Brillo1}) through a simplistic argument based on Einstein's formula $\overline{\,(x-x_0)^2\,} = 2Dt$ for the mean squared displacement of a diffusing particle in time $t$; the factor ${{\textstyle\frac{1}{2}}}$ was introduced to account for the fact that half of the particles move towards the wall, and the other half in the opposite direction. We note that our $M$ corresponds to Brillouin's symbol ${\mathfrak N}$.\par

\begin{figure}[!h]
\begin{minipage}[b]{0.475\linewidth}
\centering
\includegraphics[width=\textwidth]{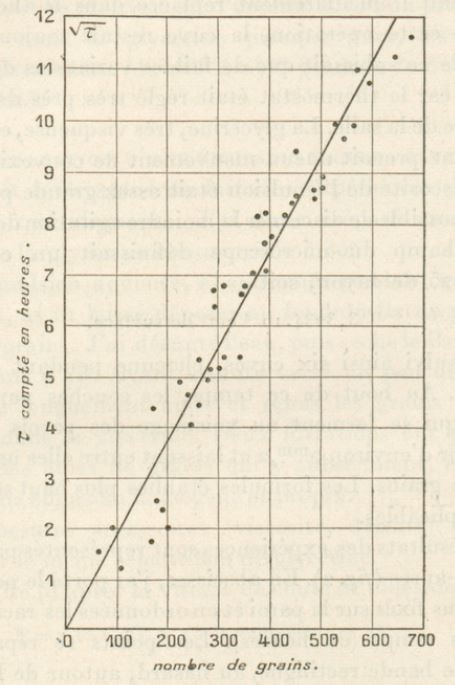}
\caption{Figure 2 of Brillouin's article \cite{Brillouin1912AnnChimPhys}; see text for more details.}
\label{fig:figure1}
\end{minipage}
\hspace{0.8cm}
\begin{minipage}[b]{0.45\linewidth}
\centering
\includegraphics[width=1.033\textwidth]{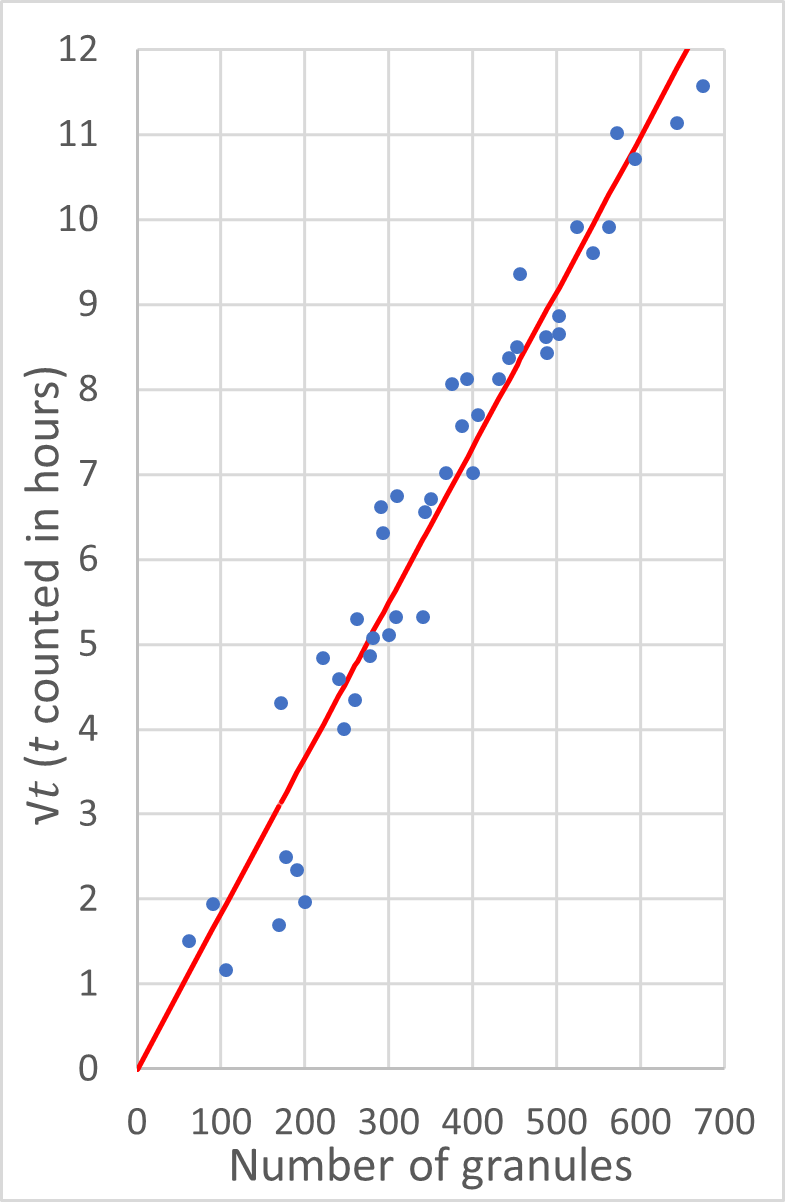}
\caption{``Reconstruction'' of Figure \ref{fig:figure1}; see text for more details.}
\label{fig:figure2}
\end{minipage}
\end{figure}

Brillouin plotted $\sqrt{t}$ against $M$ and found that the data fitted the straight line $M=54\sqrt{t}$ (see Fig. \ref{fig:figure1}). Using this value of the slope and Eq. (\ref{eq:Brillo1}), Brillouin was led to the result $D=2.32\times 10^{-11}\ {\rm cm}^2\ {\rm s}^{-1}$, which yielded, when inserted in Eq. (\ref{eq:StokesEinst}) $N_{\rm A}=6.9\times 10^{23}$, in excellent agreement with other determinations in Perrin's laboratory.\par

Since Brillouin's data points will be needed later for reproducing a plot made by Burger, I have replotted the data (Fig. \ref{fig:figure2}), and displayed the result alongside an image of Brillouin's own plot to facilitate a visual comparison. The data points from his graph were read off visually, but the reading errors are believed to be too small to vitiate the analysis presented here; it need hardly be added that the visually retrieved data points in Fig. \ref{fig:figure2} conform to the relation $M=54\sqrt{t}$.\par

\end{multicols}
\vspace{-30pt}
\setlength{\unitlength}{1mm}
\begin{picture}(100, 10)(0,0)
\put(-7, 0){\line(1,0){183}}
\put(-7, 0){\line(0,1){2}}
\put(176, 0){\line(0,1){2}}
\color{black}
\end{picture}
\vspace{-2mm}

\vspace{-15mm}
\setlength{\unitlength}{1mm}
\begin{picture}(100, 10)(0,0)
\put(-7, 0){\line(1,0){183}}
\put(-7, -2){\line(0,1){2}}
\put(176, -2){\line(0,1){2}}
\color{black}
\end{picture}

\begin{multicols}{2}

\section{Three Lectures of Smoluchowski}
\label{sec:ThreeLectures}
MvS discussed Brillouin's data on three occasions: once in a 1915 paper whose title ends with the words ``Brillouin's Diffusionversuche'' \cite{Smoluchowski1915AkadWien}, and twice (pp. 569--71, 585--86) in a three-part compendium of his results on diffusion, Brownian motion and colloidal coagulation, printed in two instalments \cite{Smoluchowski1916PhysZeit}. I will quote from the compendium (hereafter called \textit{Drei Vorträge}, meaning ``three lectures''), partly because Burger refers only to this work, and partly because an English version may be found in a recently publised selection of MvS's works \cite[pp. 125--74]{Cichocki2017Smoluchowski}. The free translations given below are my own efforts. In what follows, I will refer to Eq. (\#) of \textit{Drei Vorgträge} as Eq. ({\textit{DV}-\#}).

In \S2 of Part I, Smoluchowski models Brownian motion in the presence of an absorbing boundary as a random walk on a regular lattice with an absrobing barrier, and uses purely combinatorial reasoning---that is, without appealing to Fick's law or the DE---to arrive at the following relation:
\vspace{-6pt}
\begin{equation}\label{eq:DV-39}
M=n_0\,\hspace{-2pt}\int\displaylimits_0^t\hspace{-4pt} \sqrt{\frac{D}{\pi t}}\d t = \frac{2n_0\sqrt{Dt}}{\sqrt{\pi}}. \tag{\textit{DV}-39}
\vspace{-6pt}
\end{equation}

MvS compares the above result with Eq. (\ref{eq:Brillo1}) of Brillouin and notes that it\par
\medskip
\leftskip=20truept
\setstretch{0.9}
{\small\noindent
gives a result that is $2 \sqrt{2/\pi}$ larger than Brillouin's calculation. His experiments imply a value of $D$ that is too small by a factor $\pi/8$ and, accordingly, a completely wrong value for the Loschmidt number ($N = 176\cdot 10^{22}$). This method seems to suffer from some fundamental flaw.\par

A simple explanation of this discrepancy presents itself to me: the assumption that every particle that hits the wall instantly sticks to it did not hold in those experiments. A  direct experimental verification is ruled out, since it is not possible to observe whether the particles really stick to the wall at the first collision, or whether they have to collide several times on average before they stick.\par

If the latter is the case, the observed reduction in the number of adsorbed particles is readily explained. The physical reason for sticking, judging from electrosmotic phenomena, is likely to be that the electrical double layer on the surface of the wall and the particles, which otherwise acts like an elastic cushion, disappears as a result of the addition of glycerine or electrolyte, so that the capillary Attractive forces come into effect (or perhaps even special attractive forces arise). It would therefore be expected that immediate sticking would occur if the wall was completely discharged, as occurs when a large amount of electrolyte is added; but whether this was the case in those experiments is completely unknown.\par

Notwithstanding its ingenuity, the method does not seem to be conducive for obtaining reliable results.}\par
\leftskip=0truept
\setstretch{1.04}

In \S3 (Part II, p. 585), MvS continues with Brillouin's problem and points to a simpler approach:\par
\leftskip=20truept
\setstretch{0.9}
{\small\noindent
Now we want to turn our attention to the mathematical side of the investigation. Is it not possible to apply the \textit{macroscopic diffusion theory} here in some form instead of going back to the \textit{microscopic mechanism of Brownian displacements}? [Emphasis added here.] Evidently, this only requires a mathematical formulation of the absorbing property of the wall, and the boundary  condition that the particle concentration vanishes on the wall will be sufficient for this purpose$^1)$. [FN 1 is reproduced after this block quote.]\par
In fact, our earlier formulas can be easily found if one knows the solution to the problem of determining the distribution of a substance which at time $t = 0$ from $x = 0$ to $x = \infty$ had the uniform concentration $n = n_0$, and which  satisfies, in addition to the differential equation for diffusion:
\vspace{-4pt}
\begin{equation}
\frac{\partial n}{\partial t}=D\frac{\partial^2 n}{\partial x^2} \tag{\textit{DV}-40}
\vspace{-4pt}
\end{equation}
the boundary condition $n=0$ for $x=0$ at all times $t>0$.}\par
\leftskip=0truept
\setstretch{1.04}

We recall the text of footnote 1:  ``Since the `speed' of Brownian motion is infinitely large for infinitely small distances, the adsorbing property of the wall must cause a complete removal of the particles from an infinitely thin layer adjacent to it.'' Smoluchowski shows no hesitation in going to the continuum limit (infinitely small diffusive displacements, each covered at infinite speed); in fact, going to this limit would also relieve him of the burden of paying special consideation to a spherical absorbing surface. 

A closer look leads one to the following conclusion about the continuum limit \cite{NGvK+Oppenheim1972MasterEq,KRN1982CPL-RW1,KRN1982CPL-RW2}: it facilitates the computation, but it obliterates the distinction between a perfect absorber ($\varepsilon=1$) and an almost perfect reflector ($\varepsilon$ arbitrarily small but different from zero)! \textit{In the continuum limit, the particle concentration vanishes at any boundary that is not a perfect reflector!}

MvS finds the particle concentration for the above problem to be 
\vspace{-4pt}
\begin{equation}
n=\frac{2\,n_0}{\sqrt{\pi}} \int\displaylimits_0^{\textstyle{\frac{x}{2\sqrt{Dt}}}} \e^{-y^2}\, \mathrm{d} y, \tag{\textit{DV}-41}
\vspace{-4pt}
\end{equation}
and uses it to calculate the flux into the wall, which comes out to be
\begin{equation}
\textsf{N} \d t=D \left.\frac{\partial n}{\partial x}\right |_{\raisebox{-6pt}{\hspace{-6pt}\scriptsize $x=0$}}\hspace{-6pt}\d t =n_0\sqrt{\frac{D}{\pi t}} \d t, \tag{\textit{DV}-41+}
\end{equation}
in agreement with Eq. (\textit{DV}-38), which was obtained by an argument not involving Fick's law.

MvS sums up his theory of colloidal adhesion to a plane wall in the following words:\par
\smallskip
\leftskip=20truept
\setstretch{0.9}
{\small\noindent
Thus \ldots the average effect resulting from the Brownian motions of the individual particles, together with the adsorbing effect of the solid wall, can be treated by diffusion theory if one introduces the boundary condition that the concentration of particles on the solid wall is always zero. This seems very plausible when we consider the mechanism of diffusion, and we shall be entitled to apply the same method of calculation to an absorbing surface of arbitrary shape of the wall.\par

We will mention here a certain application of those considerations, which will turn out to be useful in the development of the theory of coagulation theory. Let us formulate the problem of calculating in a similar manner the number of particles which would adhere to a perfectly adsorbing spherical surface of radius $R$ \ldots.
}\par
\leftskip=0truept
\setstretch{1.04}
\medskip

What MvS wrote absorption by spheres will be briefly recalled in \S\ref{subsec:BurgerSphere}.

\section{Burger's ingenious approach for dealing with absorbing boundaries}
\label{sec:BurgAnal}
In what follows, Eq. (\#) of HCB1918 will be labelled as Eq. (B-\#). \S\S 1 and 2 of Burger's paper, which are devoted to the derivation of the DE, will not be reviewed here. 
The last numbered equation in \S2 is Eq. (B-\textit{X}). Burger's analysis of Brillouin's data begins with Eq. (\ref{eq:B-17}), but we must first become familiar with two functions introduced by Burger at the start of \S2:
\begin{enumerate}[wide, labelwidth=!, labelindent=0pt]
\item The probability, $f(x,p,t)\d p$, that a particle, known to be at the abcissa $x$ at $t=0$, will have the abcissa between $p$ and $p+\d p$ at time $t$.
\item The probability $\chi(x,t)$ that the particle has been adsorbed during the interval $t$.
\end{enumerate}

Since a particle that is free (not adsorbed) at the time $t_1+t_2$ was free also at $t_1$, $f$ will satisfy the integral equation 
\vspace{-4pt}
\begin{equation}\label{eq:B-11}
\int_0^\infty \hspace{-4pt} f(x,p,t_1)f(p,a,t_2)\d p=f(x,a,t_1+t_2). \tag{B-\textit{XI}}
\vspace{-1pt}
\end{equation}
A similar argument leads Burger to the following integral equation for $\chi$:
\begin{equation}\label{eq:B-12}
\chi(x,t_1+t_2)=\chi(x,t_1) +
\int_0^\infty \hspace{-4pt} f(x,p,t_1)\chi(p,t_2)\d p. \tag{B-\textit{XII}}
\end{equation}
Since the particle must be somewhere at time $t$, a thrid relation immediately follows:
\begin{equation}\label{eq:B-13}
\int_0^\infty \hspace{-4pt} f(x,p,t)\d p +\chi(x,t)=1. \tag{B-\textit{XIII}}
\vspace{-4pt}
\end{equation}

For the system studied by Brillouin (no applied field, $x=0$ an absorbing boundary), $f(x,p,t)$ may be found by solving the diffusion equation
\vspace{-4pt}
\begin{equation}\label{eq:B-14}
\frac{\partial f}{\partial t}= D \frac{\partial^2 f}{\partial p^2}, \tag{B-\textit{XIV}}
\vspace{-4pt}
\end{equation}
but a boundary condition remains to be found. Burger manipulates Eq. (\ref{eq:B-13}) to get the differential equation
\begin{equation}\label{eq:B-15}
\frac{\partial \chi(x,t_1)}{\partial t_1}=\varkappa f(x,0,t), \tag{B-\textit{XV}}
\vspace{-4pt}
\end{equation}
where $\varkappa$ is a constant (that cannot be determined without additional input).

Burger integrates Eq. (\ref{eq:B-14}) to get
\begin{equation}\label{eq:B-13}
\frac{\partial }{\partial t}\int_0^\infty \hspace{-4pt} f(x,p,t)\d p = - D\left(\frac{\partial f}{\partial p}\right)_{p=0}, \tag{B-\textit{XIII}}
\vspace{-4pt}
\end{equation}
and uses Eqs. \ref{eq:B-13} and \ref{eq:B-15} to arrive at the boundary condition
\vspace{-4pt}
\begin{equation}\label{eq:B-16}
D\left(\frac{\partial f}{\partial p}\right)_{p=0} = \varkappa f_{p=0}, \tag{B-\textit{XVI}}
\vspace{-4pt}
\end{equation}
and points out that if one's interest lies in finding the number of particles attached to the wall, one need not work with $f$, because the initial condition for this function ``is less simple''. Accordingly, he introduces a new function
\vspace{-4pt}
\begin{equation}\label{eq:B-17}
F(p,t)=n_0\int_{0}^{\infty}\hspace{-4pt} f(x,p,t) \d x.\quad {}^1) \tag{B-\textit{XVII}}
\vspace{-4pt}
\end{equation}
Footnote 1, indicated on the right-hand side of Eq. (\ref{eq:B-17}), explains that ``$F(p,t)$ is the concentration of the particles at the time $t$ at a distance $p$ from the wall.'' That is to say, our $n(x,t)$ is Burger's $F(p,t)$. Burger continues:\par
\medskip
\leftskip=20truept
\setstretch{0.9}
{\small\noindent
Integrating (\textit{XIV}) and (\textit{XVI}) with respect to $x$ from zero to infinity, it is easily seen that $F(p,t)$ satisfies the same differential equation and boundary-condition as $f(x,p,t)$, considered as a function of $p$ and $t$. Further it follows from (\textit{XIII}) and (\textit{XVII}) that the beginning [initial] condition is $F(p,0)=n_0$.\par
The solution of the differential-equation for $F$, regarding the beginning and boundary condition is
\[
F(p,t)= n_0  \ldots \ldots \ldots \ldots \ldots \ldots   {}^2) \tag{B-\textit{XVIII}}
\]
}\par
\leftskip=0truept
\setstretch{1.04}
\noindent

The footnote indicated on the right-hand side of the above equation is a bibliographic item, referring to H. Weber's authoritative work on partial differential equations \cite{Weber1912Partiellen}. Carslaw's first book on heat conduction, cited by C\&K, did not appear until 1922.

Burger's result for $M$ ($n_t$ in his notation) will be restated in a more compact form as follows:
\vspace{-4pt}
\begin{equation}\label{eq:B-19}
M = 2n_0\sqrt{\frac{Dt}{\pi}}+\frac{n_0D}{\varkappa}\left [e^{{\pmb\alpha}^2t}\mbox{erfc}({\pmb\alpha}\sqrt{t})-1\right ],  \tag{B-\textit{XIX}}
\end{equation}
where ${\pmb\alpha}=\varkappa/D$ [Burger's $\alpha$ has been replaced by ${\pmb\alpha}$, because $\alpha $ is frequently used for the sticking probability (MvS's $\varepsilon$)]. The first term on the right-hand side of Eq. (\ref{eq:B-19}) coincides with the right-hand side of Eq. (\ref{eq:DV-39}).

After presenting Eq. (\ref{eq:B-19}), Burger prepares the background for discussing its application to Brillouin's data:\par
\medskip
\leftskip=20truept
\setstretch{0.9}
{\small\noindent
With this the problem is solved that is mentioned by v. Smoluchowski with respect to the fact that this [his?] tbeory was not in accordance with the experimental results of Brillouin. The latter has experimented with particles of gamboge in a mixture of glycerine and water. The number of the particles that was adsorbed by the wall could be determined by counting them on a microfotograph.\par 
Now, while Brillouin has concluded, that his observations agree with theory, v. Smoluchowski has pointed out the incorrectness of this method of reckoning and has substituted this by a better one. Thereby he supposes however that every particle that collides against the wall, sticks to it \ldots. The result of this computation agrees very badly with the observations and v. Smoluchowski thinks that this is a consequence of the fact, that a particle, colliding against the wall, does not adhere at once, but on the average has to collide several times befor'e being adsorbed. So this would mean, that the boundary condition has to be altered.\par

\hspace{-3pt}As now is demonstrated by \mbox{v. Smoluchowski}, we find the solution of the problem as given by him, from the diffusion equation, when we use $F(0,t) = 0$ as boundary condition, while (\textit{XIX}) is deduced with the general limiting-condition [boundary condition]:
\vspace{-4pt}
\begin{equation}\label{eq:ABC}
D\left(\frac{\partial F}{\partial p}\right)_{p=0} = \varkappa F_{p=0}. \tag{B-p654}
\vspace{-4pt}
\end{equation}

So we may expect that by a suitable choice of $\varkappa$ we get a result that is more in accordance with the experiments of Brillouin.\par

To be able to decide in how far this is the case we compute from the data furnished by Brillouin, $n_0$ and $D$ the latter with the help of of (Ib) [Eq. (\ref{eq:StokesEinst})]. When we now choose a certain value of $\varkappa$, we can represent $n_t$ [our $M$] graphically as a function of $t$. This is to be seen in fig. 1 [recreated in 
Fig. \ref{fig:BurgFig1}], in which the abscissa represents the number $n_t$ of the adsorbed particles, and the ordinate, in accordance with Brillouin, the square root of the time, expressed in hours. Further, for practical reasons, not the values of $\varkappa$ but those of $\alpha=\varkappa/D$ have been written at the curves.}\par
\leftskip=0truept
\setstretch{1.04}
\bigskip
\begingroup
    \centering
    \includegraphics[width=0.5\textwidth,angle=0]{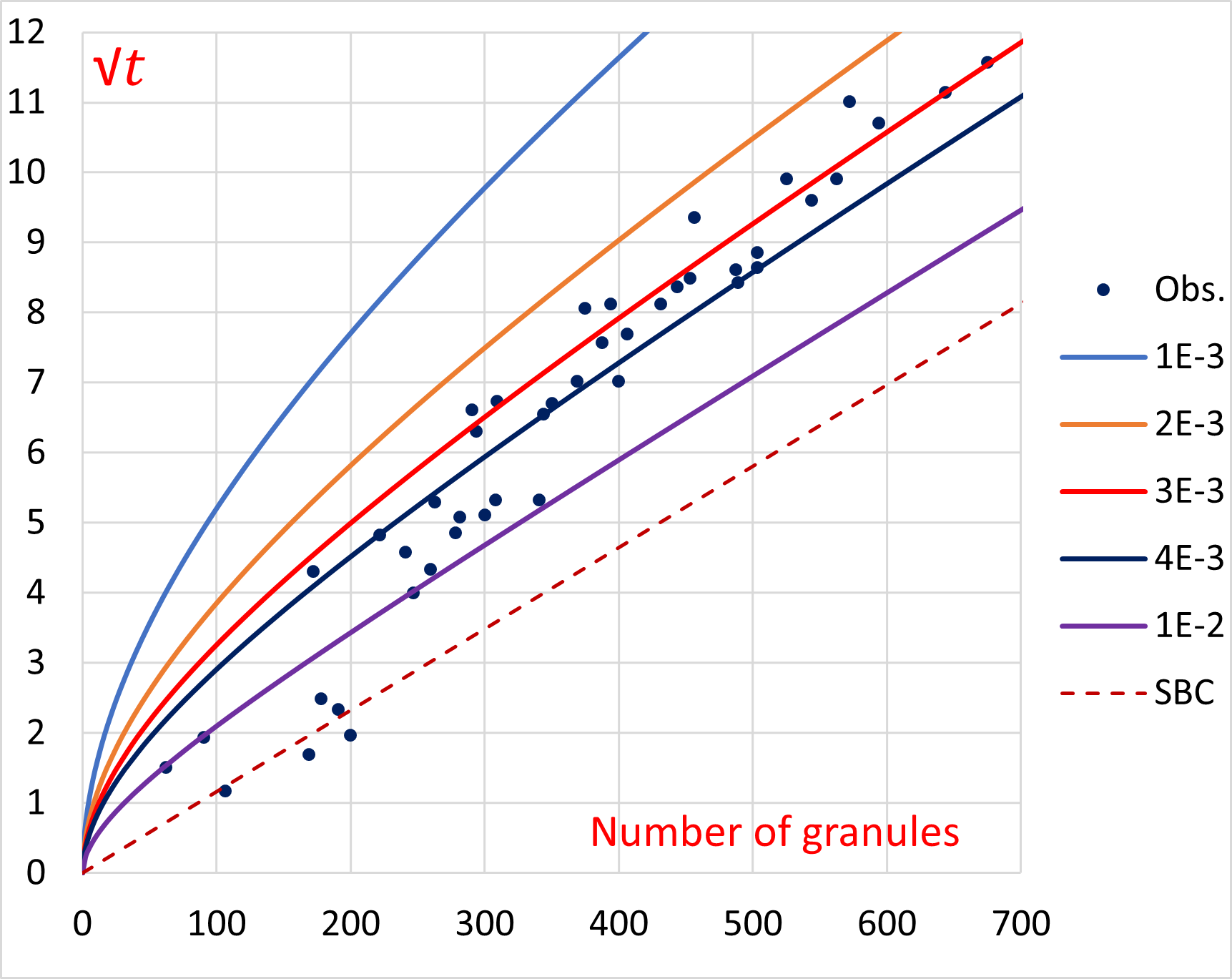}
    \captionof{figure}{\small A recreation of Figure 1 of Burger's 1918 paper. The plot shows a family of curves representing the value of $M$ accoring to Eq. (\ref{eq:B-19}) with the following values of ${\pmb\alpha}=\varkappa/D$: 0.001, 0.002, 0.003, 0.004, 0.01; for other details, see text. The dashed straight line corresponds to ${\pmb\alpha}=\infty$.  }
\label{fig:BurgFig1}
\endgroup
\bigskip
We hand over the discussion of the plot to Burger himself:\par
\medskip
\leftskip=20truept
\setstretch{0.9}
{\small\noindent
For $\varkappa= \infty$  $({\pmb\alpha} = \infty)$ we get a straight line, agreeing with the theory of v. Smoluchowski. One sees that the observations of Brillouin, indicated by crosses [here filled circles], do not at all correspspond with them. \par
Smaller values of $\varkappa$ or $\alpha$ give  curves that agree better with the observations, though none of the curves gives an entirely satisfying result. Probably  ${\pmb\alpha}=0.003$ $(\varkappa=1.5 \times 10^{-8})$ comes nearest to the truth, when we bear in mind, that the observations made after short times are least to be trusted. Then the influence of small differences of temperature may viz, still be observed; this influence disappears after a long time. Also, because we can only determine theoretically
the least probable number, the observations after a long time, which deliver larger numbers, have more importance.\par

It is not to be denied however, that there seems to be a systematic error. Perhaps this may be sought in the fact of the particles not being exactly equal, or in a slow change of the electrical double-layers at the particles and the wall, in consequence of the numerous collisions. Further observations of the behaviour of colloidal particles with respect to a fixed wall, will have to decide whether this explanation is tenable.\par

A remarkable circumstance is, that the curves practically coincide with the straight line for $\varkappa= \infty$ $({\pmb\alpha} = \infty)$ when $\varkappa$ or ${\pmb\alpha}$ are above a certain small value. With regard to the precision of the measurements all cases in which    $\varkappa> 5 \times 10^{-7} (\alpha > 0.01)$ may be treated
as if $\varkappa= \infty$ $({\pmb\alpha} = \infty)$.  If the observations agree with the theory for $\varkappa= \infty$, we may consequently only conclude from this that $\varkappa$ is larger than this small value.}\par

\leftskip=0truept

\setstretch{1.04}

\subsection{Significance of the proportionality constant in Burger's alternative boundary condition}
\label{subsec:SigProp}

Having concluded that the rate of absorption of the particles was proportional to the concentration of the particles at the wall, Burger next task was to establish the physical significance of the proportionality constant $\varkappa$:\par
\medskip
\leftskip=20truept
\setstretch{0.9}
{\small\noindent
The quantity $\varkappa$ occurring above has the dimension of a velocity and to know what the values of $\varkappa$ mentioned before mean, we have to find out with what velocity $\varkappa$ has to be compared. So we have to enter into a closer investigation of the physical meaning of $\varkappa$, also to decide whether $\varkappa$ may assume all values between zero and infinity as has been tacitly supposed up till now.\par

From the way in which (\textit{XIX}) is deduced, follows [Our $M$ is Burger's $n_t$]:
\vspace{-4pt}
\begin{equation}\label{eq:B-20}
\frac{\d n_t}{\d t} = \varkappa F(0,t)\tag{B-\textit{XX}}
\vspace{-4pt}
\end{equation}

Now bearing in mind, that $\d M/\d t$ represents the number of the particles that sticks to the wall in a unit of time, it naturally suggests itself to compare this with the number of the particles that collides against the wall. This number, as is known from the kinetic theory of gases and liquids [see below], amounts to:
\vspace{-4pt}
\begin{equation}\label{eq:B-21}
\frac{v}{\sqrt{6\pi}}F(0,t) = \varkappa_1 F(0,t),		\tag{B-\textit{XXI}}
\vspace{-4pt}
\end{equation}
where $v$ is the mean velocity of the particles that is given by the known formula:
\vspace{-4pt}
\begin{equation}\label{eq:B-22}
v=\sqrt{\frac{3RT}{N_{\rm A} m}} \tag{B-\textit{XXII}}
\vspace{-4pt}
\end{equation}
}\par
\leftskip=0truept
\setstretch{1.04}
I have interrupted Burger in order to make some comments that must not be postponed. 

\textbf{First}: The symbol $\varkappa_1$ on the right-hand side of Eq. (\ref{eq:B-21}) has not been defined yet; its significance is explained in Eq. (\ref{eq:B-23}):
\vspace{-4pt}
\begin{equation}\label{eq:B-23}
\varepsilon=\frac{\varkappa\,F(0,t)}{\varkappa_1 F(0,t)}=\frac{\varkappa}{\varkappa_1}  \tag{B-\textit{XXIII}}.
\vspace{-4pt}
\end{equation}

\textbf{Second}: Eq. (\ref{eq:B-21}) ends with a pointer to a footnote, which reads:\par
\medskip
\leftskip=20truept
\setstretch{0.9}
{\small\noindent When we want to take the change of the concentration with the distance from the wall into account, we have to substitute $F_{p=0}+\lambda (\partial F/\partial p)_{p=0}=0$ for $F_{p=0}=0$, where $\lambda$ is a length of the order of the mean free path. [Each equation should be changed into an expression by deleting $=0$ at the end.] As $D(\partial F/\partial p)_{p=0}=\varkappa F_{p=0}$, the ratio of the correction to the used value is $\varkappa \lambda/D$, i.e. always very small.}\par
\medskip
\leftskip=0truept
\setstretch{1.04}

It is necessary to observe that, since the right-hand side of Eq. (\ref{eq:B-21}) is $\varkappa_1$, it can only be applied to a black wall ($\varepsilon=1$). Accordingly, the symbol $\varkappa$ in his footnote should be replaced by $\varkappa_1$; alternatively, the equation should be made applicable to an elastic wall ($0\leq \varepsilon\leq 1$) by multiplying both sides by $\varepsilon$.\par
\textbf{Third}: The right-hand side of Eq. (\ref{eq:B-22}) is in fact the root-mean-squared velocity $[\overline{v^2}]^{1/2}$, which will be represented here by the symbol $\hat{v}$. The left-hand side of  Eq. (\ref{eq:B-21}) can be identified with $\overline{v}/4$, where $\overline{v}=(8\,k_{\raisebox{-1pt}{\scriptsize B}}T/\pi m)^{1/2}$ is the average speed (often called the average velocity).

To prepare the path for retracing the logical steps taken by Burger in deducing Eqs. (\ref{eq:ABC}), (\ref{eq:B-20})--(\ref{eq:B-23}), all these relations will be distilled into a single set of four equalities (stated in our notation):
\vspace{-4pt}
\begin{align}
\frac{\d M}{\d t}&= D \left. \frac{\partial n}{\partial x}\right |_{x=0}= \varkappa n(0,t)\\
&= \varepsilon\varkappa_1 n(0,t)=\frac{\varepsilon\overline{v}}{4}n(0,t)
\end{align}

Since the partial current $j{\hspace{-2pt}\smash[t]{\strut}}_{-}(0,t)$ represents the rate at which the suspended particles impact unit area of the wall, the rate of particle absorption by an elastic wall ($0\leq\varepsilon\leq 1$) will equal $\varepsilon j{\hspace{-2pt}\smash[t]{\strut}}_{-}(0,t)$.
 Burger proceeds by applying (but not explicitly stating) the relation
 \vspace{-4pt}
\begin{equation}\label{eq:D-as-Sum-jpm}
D\left( \frac{\partial n}{\partial x}\right)_{x=0}= \varepsilon j{\hspace{-2pt}\smash[t]{\strut}}_{-}(0,t) 
\vspace{-4pt}
\end{equation}
to the wall at $x=0$, and goes through the following steps:
\begin{align}
D \left. \frac{\partial n}{\partial x}\right |_{x=0}
&=\frac{\varepsilon\overline{v}}{4}\left[n(x,t) + \lambda \frac{\partial n}{\partial x}\right]_{x=0} \label{eq:Jm2terms}\\
\noalign{\vspace{4pt}}
&={\textstyle\frac{1}{4}}\varepsilon\overline{v}n(0,t). \label{eq:fallacy}
\end{align}
Burger justifies the last step by claiming, in the footnote, that $\lambda (\partial n/\partial x)_{x=0}\ll n(0,t)$, where $\lambda$ is approximately equal to the mean free path (mfp). Although the exact relation between $\lambda$ and mfp is not devoid of interest, we need not go into that detail here. Let us simply note that the partial currents across an imaginary plane (located at abcissa $x$) perpendicular to the $x$-axis are given by the relations
\vspace{-6pt}
\begin{equation}\label{eq:Jpm}
j{\hspace{-2pt}\smash[t]{\strut}}_{\pm}(x,t)=  \frac{\overline{v}}{4}\left[n(x,t) \mp \lambda \frac{\partial n}{\partial x}\right],
\vspace{-6pt}
\end{equation}
which lead us immediately to Fick's law,
\vspace{-6pt}
\begin{equation}\label{eq:NetFlux}
j(x,t)=j{\hspace{-2pt}\smash[t]{\strut}}_{+}(x,t)-j{\hspace{-2pt}\smash[t]{\strut}}_{-}(x,t) =-\frac{\overline{v}\lambda}{2}\frac{\partial n}{\partial x}
\vspace{-6pt}
\end{equation}
with the diffusion coefficient given by the relation
\vspace{-4pt}
\begin{equation}
D=\frac{\overline{v}\lambda}{2}. \label{eq:Fick-lambda}
\vspace{-6pt}
\end{equation}

We can see now that, $\frac{1}{4}\overline{v}\varepsilon\lambda (\partial n/\partial x)_{x=0}$, the second term on the expanded right-hand side of Eq. (\ref{eq:Jm2terms}), equals $\frac{1}{2}\varepsilon D(\partial n/\partial x)_{x=0}$, which would become, when $\varepsilon$ is close to unity, \textit{comparable to the left-hand side}; at the opposite extreme, $\varepsilon\ll 1$, the second term can indeed be neglected, and Burger's boundary condition would then become defensible.  With this in mind, we return to Eq. (\ref{eq:Jm2terms}), expand its right-hand side, and express the second term in a form suitable for transposition:
\vspace{-4pt}
\begin{equation}\label{eq:JmBoth}
D\left( \frac{\partial n}{\partial x}\right)_{x=0}=\frac{\varepsilon\overline{v}}{4}n(0,t) + \frac{\varepsilon D}{2} \left( \frac{\partial n}{\partial x}\right)_{x=0}.
\vspace{-4pt}
\end{equation}
The last equation can be easily rearranged to get a fallacy-free version of Eq. (\ref{eq:fallacy}):
\vspace{-4pt}
\begin{equation}\label{eq:TBC}
D\left( \frac{\partial n}{\partial x}\right)_{x=0}=\frac{\varepsilon}{2-\varepsilon}\frac{\overline{v}}{2}n(0,t) ,
\vspace{-4pt}
\end{equation}
an equation that I propose to baptize as the Trondheim boundary condition (TBC), because the Trondheim group has derived this BC in more than one way and for the reduction of more than on transport equation, often with the aid of the relation
\vspace{-2pt}
\begin{equation}\label{eq:Flux=Jm}
j{\hspace{-2pt}\smash[t]{\strut}}_{+}(0,t) = \kappa j{\hspace{-2pt}\smash[t]{\strut}}_{-}(0,t),\quad (\kappa\equiv 1-\varepsilon),
\vspace{-2pt}
\end{equation}
instead of Eq. (\ref{eq:D-as-Sum-jpm}). When applied to a black wall, Eq. (\ref{eq:Flux=Jm}) may be stated as
\vspace{-2pt}
\begin{equation}\label{eq:BlackWall}
j{\hspace{-2pt}\smash[t]{\strut}}_{+}(0,t) = 0\;\mbox{\textit{at a perfectly absorbing surface.}}
\vspace{-2pt}
\end{equation}

In going from Eq. (\ref{eq:Jm2terms}) to Eq. (\ref{eq:fallacy}), Burger succumbed to an error that has tripped many later authors, including Noyes \cite{Noyes1954JCP}, Monchick \cite{Monchick1956JCP}, Zhou and Szabo \cite{ZhouSzabo1991JCP}, and the present author \cite{KRN+SW+KJM1979JCP}. 

As already noted, Burger's BC, Eq. (\ref{eq:fallacy}), is a special case of TBC, Eq. (\ref{eq:TBC}). 
When the sticking factor $\varepsilon$ is much smaller than unity (a highly reflecting surface), the latter boundary condition becomes practically indistinguishable from Burger's BC, which stands to reason, because Burger ignored a term containing $\partial n/\partial x$, and the derivative will actually vanish in the case of a perfectly reflecting wall, whose presence does not distrub the equilibrium. Departure from equilibrium, negligible when $\varepsilon\ll 1$, becomes increasingly pronounced as the wall becomes more absorbing, and Eq. (\ref{eq:fallacy}) loses its validity.

\subsection{Absorption by a sphere}
\label{subsec:BurgerSphere}
In the last and the shortest section of his paper, Burger remarks that though the boundary condition in \S3        
 is deduced explicitly for a plane wall, ``we may undoubtedly also apply it to a curved wall
in the form:
\begin{equation}\label{eq:B-24}
D\frac{\partial \overline{F}}{\partial \nu}= \varkappa \overline{F} \tag{B-\textit{XXIV}}
\end{equation}
where the horizontal line denotes values at the wall and $\nu$ is the normal directed towards the liquid.''\par

Burger continues:\par
\medskip
\leftskip=20truept
\setstretch{0.9}
{\small\noindent
This we may apply to the case of a fixed sphere with a radius $R$, surrounded by a liquid in which at the time $t=0$ many particles are dispersed homogeneously ($n_0$ per unit of volume). 

The solution of this problem is namely used by v. Smoluchowski in his theory of the coagulation. Where \textit{he} however uses as boundary-condition
$\overline{F}= 0$, we will here take (\textit{XXIV}) as limiting-condition. The number of particles that sticks in a unit of time after some computations
proves to be:
\begin{align}
\frac{\d n_t}{\d t}&=4\pi RD n_0 \frac{\varkappa R}{\varkappa R + D}\nonumber\\
&=4\pi RD n_0\frac{\varepsilon\varkappa_1 R}{\varepsilon\varkappa_1 R + D}\tag{B-\textit{XXV}}\label{eq:B-25}
\end{align}
when we, like v. Smoluchowski, only consider times that are large with respect to $R^2/D$.
}\par
\medskip
\leftskip=0truept
\setstretch{1.04}

Burger devotes the last paragraph to a comparison of the above result with the corresponding expression of MvS:\par
\medskip
\leftskip=20truept
\setstretch{0.9}
{\small\noindent
When $\varepsilon = 1$ or is at least not too small, we get very approximately $\varepsilon\varkappa_1 R/(\varepsilon\varkappa_1 R + D)=1$
and (\textit{XXV}) becomes:
\begin{equation}\label{eq:B-25a}
\frac{\d n_t}{\d t}=4\pi RD n_0, \tag{B-\textit{XXV}$^a$}
\end{equation}
as is also found by v. Smoluchowski. If $\varepsilon$ is however very small, then the same formulae remain valid as is also remarked by v. S., when we only multiply $t$ with a constant fartor. This factor however is not, as v. S. assumes $\varepsilon$, but $\varepsilon\varkappa_1 R/(\varepsilon\varkappa_1 R + D)$. So, as long as $\varepsilon$ does not become very small, the formula (\ref{eq:B-25a}) of v. Smoluchowski holds good. When however $\varepsilon$ beromes a very small fraction, then the number of particles that sticks [sic] per unit of time also becomes smaller, so that in this way the slow coagulation may be explained.
}\par
\medskip
\leftskip=0truept
\setstretch{1.04}

Burger is referring, in the last sentence, to a problem mentioned by Smoluchowski in \S6 (p. 597) of the final part of the \textit{Drei Vorträge}:\par
\medskip
\leftskip=20truept
\setstretch{0.9}
{\small\noindent Finally, I would like to briefly remark that our theory seems applicable, at least from a formal aspect, to the phenomena of ``slow'' coagulation, which occur when very little electrolyte is added. To this end, it would be sufficient to introduce the plausible assumption that under these circumstances, i.e., when the electrical discharge of the particles is incomplete, only a certain fraction $\varepsilon$ of the wall-particle collisions leads to their capture. As to how large this fraction is, we know nothing in advance, except that it depends to a large extent on the charge of the double layer; but we will see that it can be determined ``a posteriori'' from the experimental data.\par
Under this assumption [Unter dieser Annahme], it becomes evident that formulas (69, 70) would remain valid also in this general case, with the only difference that the term $\beta t$ is to be replaced by $\varepsilon\beta t$ everywhere [$\beta$, the stationarynvalue of the flux, equals $4\pi RDn_0$, see Eq. \ref{eq:B-25a} above].  Thereby follows without further ado the important statement [\textit{Es folgt also ohne weiteres der wichtige Satz}] that the coagulation curves obtained at different concentrations of colloid and electrolyte must be similar, in the sense that they can be made to coincide by an appropriate change of the time scale. The times required to achieve a certain degree of coagulation are therefore inversely proportional to the product ($\varepsilon \nu_0$).}\par
\medskip
\leftskip=0truept
\setstretch{1.04}
Considering the importance of the topic, one only wishes that there had been much further ado, that MvS had favoured his readers with a hint or two, if not a detailed explanation as to how he reached this conclusion, instead of leaving the task to his expositors, who---with the exception of Fuchs---were content to reproduce what they found in MvS's writings.

\subsection{Fuch's 1934 paper on the theory of coagulation}
\label{eq:Kolmogorov}

Fuchs made several comments on Smoluchowski's theory of coagulation in a 1934 article \cite{Fuchs+Y1934+199}, entitled ``Zur theorie der koagulation''. A non-verbatim English rendering of the bulk of the abstract, is given below: \par
\medskip
\leftskip=20truept
\setstretch{0.9}
{\small\noindent 
The basic equation of Smoluchowski's theory of ``fast coagulation'' is reformulated, using the method of Kolmogoroff \ldots  the theory does need an important correction whose magnitude depends on the value of the ratio $\lambda/a$, where $a$ denotes the particle radius, and $\lambda$ the ``average path length of the particle in a specific direction''. The correction, inconsequential in colloidal solutions, is far from negligible in aerosols. On the other hand, Smoluchowski's theory of ``slow coagulation'' needs to be overhauled. The coagulation rate, set by Smoluchowski to be simply proportional to the ``effectiveness'' $\alpha$ of  individual collisions [between the diffusing particles and the absorbing sphere], becomes [according to a formula derived in this paper] less and less sensitive to $\alpha$ as $\lambda/a$ becomes smaller and smaller. This aspect must be taken into account in the interpretation of experimental results on the stabilization of disperse systems.}\par
\medskip
\leftskip=0truept
\setstretch{1.04}

It would be convenient to discuss ``the method of Kolmogoroff'' in a separate section, and deal with the rest of the passage now. Here we continue with  Fuch's comments on slow coagulation: \par
\medskip
\leftskip=20truept
\setstretch{0.9}
{\small\noindent 
So far we have dealt exclusively with the so-called ``fast coagulation", in which every collision between the particles leads to them sticking together or flowing together. We now want to move on to ``slow coagulation", where only a certain fraction $\alpha$ [$\alpha = 1/f$ in the German original] is effective from all collisions. There is an oversight here by \textsc{Smoluchowski}: he believed that the coagulation constant was simply $\alpha$ [$f$ in the original] times smaller than in the case of ``fast coagulation". In reality, this problem is a little more complicated.  It must be borne in mind that in coagulation theory the coefficient $\alpha$ plays a role equivalent to that of the accommodation coefficient, which was introduced by Knudsen in the theory of heat conduction. Therefore, with $\alpha$ values less than one, we must expect an increase in the concentration jump as it decreases when $\alpha$ grows larger.}\par
\medskip
\leftskip=0truept
\setstretch{1.04}

\subsection{Analogues of Milne's problem involving gradients of velocity and temperature}
\label{subsec:compare}

In the Brillouin problem, the particle concentration never attains a non-zero steady value, but if a source supplying a constant current of particles is placed at ``infinity'', the system would attain a time-independent concentration profile. This steady-state problem is the Brownian analogue of Milne's problem.  In the Milne problem, there exists a concentration gradient parallel to the wall at $x=0$. We will recall now two analogues of this problem from the realm of kinetic theory of gases, both well know to MvS, one involving a velocity gradient perpendicular to the wall, and another involving a temperature gradient parallel to the wall. We will assume throughout that the wall at $x=0$ is stationary, and it will be sufficient for us to consider only the velocity analogue, summarize the explanation advanced by Maxwell \cite{Maxwell1879PhilTrans}, and note that the temperature analogue was investigated by MvS \cite{Smoluchowski1898AnnPhys,Smoluchowski1898PhilMag,Smoluchowski1910AnnPhys2}.

We will examine here some excerpts from the appendix to Maxwell's 1879 article \cite{Maxwell1879PhilTrans}. Although the paper is entitled ``On stresses in rarified gases arising from inequalities of temperature'', Maxwell was content to ``to determine approximately the principal phenomena in a gas which is not very highly rarified, and in which the space variations within distances comparable to $\lambda$ are not very great''.

Maxwell considered two possible fates for molecule impinging the stationary wall, diffuse reflection or specular reflection, but he calls them absorption and reflection. 
For specular reflection, the velocity of the molecule (relative to the surface normal) is reversed, the tangential velocity remaining unchanged. An absorbing surface is defined as one which captures the incident molecules, but subsequently releases them with velocities (corresponding to the temperature of the wall) whose directions are uniformly distributed in the forward hemisphere; the molecules which return to the gas in this manner are called \textit{absorbed and evaporated}. 

Maxwell explained the difficulty of the task at hand and his reasons for adopting ``a purely statistical method'':\par
\medskip
\leftskip=20truept
{\noindent\small
We might also consider a surface on which there are a great number of minute asperities of any given form, but since in this case there is considerable difficulty in calculating the effect when the direction of rebound from the first impact is such as to lead to a second or third impact, I have preferred to treat the surface as something intermediate between a perfectly reflecting and a perfectly absorbing surface, and, in particular, to suppose that of every unit of area a portion $f$ absorbs all the incident molecules, and afterwards allows them to evaporate with velocities corresponding to those in still gas at the temperature of the solid, while a portion $1-f$ perfectly reflects all the molecules incident upon it.}\par
\medskip
\leftskip=0truept

Maxwell's analysis led him to conclude that if the gas at a finite distance from the surface is moving parallel to the surface, the gas in contact with the surface will be sliding over it with the finite velocity 
\begin{equation}\label{eq:MaxApp-69}
v=G\frac{\d v}{\d x}, \tag{M69}
\end{equation}
and his aim was to derive an approximate but useful expression for $G$. After stating Eq. (69), he went on to add a few comments:\par
\medskip
\leftskip=20truept
{\noindent\small
If, therefore, the gas at a finite distance from the surface is moving parallel to the surface, the gas in contact with the surface will be sliding over it with the finite velocity $v$, and the motion of the gas will be very nearly the same as if the stratum of depth $G$ had been removed from the solid and filled with the gas, there being now no slipping between the new surface of the solid and the gas in contact with it.\par
The coefficient $G$ was introduced by Helmholtz and Piotrowski under the name of \textit{Gleitungs-coefficient}, or coefficient of slipping. The dimensions of $G$ are those of a line, and its ratio to $l$, the mean free path of a molecule, is given by the equation
\begin{equation}\label{eq:MaxApp-70}
G=\frac{2}{3}\left(\frac{2}{f}-1 \right)l \tag{M70}
\end{equation}
The value $G = 2l$, implies $f=\textstyle\frac{1}{2}$, or the surface acts as if it were half perfectly reflecting and half perfectly absorbent. If it were wholly absorbent, $G = \textstyle\frac{2}{3}l$.
}\par
\medskip
\leftskip=0truept

We will now replace $f$ by $\varepsilon$ in Maxwell's expression for the coefficient $G$, and adisplay Eq. (\ref{eq:MaxApp-69}) along with the analogs appearing in the boundary conditions for the Milne problem obtained by using the Lorentz model of a random walk and the lattice model of Smoluchowski \cite{KRN1982CPL-RW1}: 
\begin{align}
\mbox{Maxwell:}\quad&v(0)		=\frac{2l}{3}\left(\frac{2-\varepsilon}{\varepsilon}\right)\left.\frac{\d v}{\d x}\right|_{x=0}, 			\tag{M70}\\
\mbox{Loretz:}\quad&n(0)=	\frac{2\ell}{3}\left(\frac{2-\varepsilon}{\varepsilon}\right)\left.\frac{\d n}{\d x}\right|_{x=0},  	\tag{\cite{KRN1982CPL-RW1}-30}	\label{eq:RW1-Nr30}\\
\mbox{Lattice:}\quad&	n(0)=\lambda \left(\frac{1-\varepsilon}{\varepsilon}\right)\left.\frac{\d n}{\d x}\right|_{x=0}. 		\tag{\cite{KRN1982CPL-RW1}-12}\label{eq:RW1-Nr12}
\end{align}

If we permit ourselves to indulge in the reasonable assumption that Smoluchowski must have been thoroughly familiar with the aforementioned appendix, it would be natural to ask why MvS did not choose a boundary condition of the same form as Eq. (\ref{eq:RW1-Nr30}). The only explanation that comes to my mind is that Maxwell's object was to explain a phenomenon observed only in \textit{rarefied gases}, and that of Smoluchowski to quantify an analogous phenomenon in a \textit{condensed phase}, where the fictitious mean free path is exceedingly small, small enough to justify (in Smoluchowski's opinion) embracing the continuum limit ($\ell\to 0$) and using the SBC. If this speculation is regarded as tenable, one will have to conclude that MvS overlooked (or chose to defer a scrutiny of the conundrum) that, if the wall is not a perfect reflector ($\varepsilon\neq 0$), $n(0)$ will vanish in the continuum limit ($\ell \to 0$) whatever the value of $\varepsilon$.

\subsection{The continuumm limit of Smoluchowski}
\label{eq:3BC's}

The purpose of this section is to compare the results obtained by different choices for $\varkappa$. To prepare the ground, we introduce the symbol $k$, which will be called the coagulation rate constant, through the relation
\[
\frac{1}{n_0}\frac{\d n_t}{\d t}=k.
\]
A subscript ($S$, $A$, or $T$) will be added to specify the boundary condition (SBC, ABC, TBC) used for deriving the expression for $k$. It will also be convenient to introduce two abbreviations:
\vspace{-4pt}
\begin{equation}\label{eq:DeltaXi}
\Delta\equiv\frac{2D}{\overline{v}},\quad
Q\equiv \frac{\Delta}{R}.
\vspace{-4pt}
\end{equation}

\medskip
\begingroup
   \hspace{-26pt}
    \includegraphics[width=0.5\textwidth,angle=0]{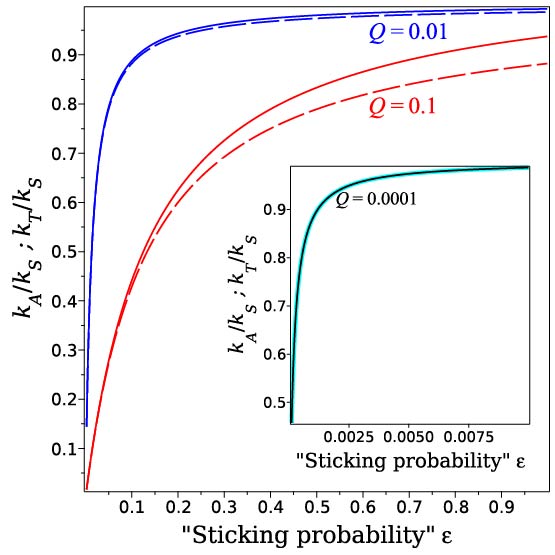}
    \captionof{figure}{\small Plots of $k_A/k_S$ [Eq. (\ref{eq:CoagConstb}), dashed curves] and $k_T/k_S$ [Eq. (\ref{eq:CoagConstc}), solid curves] against $\varepsilon$ for two values of $Q\equiv\Delta/R =0.1$ and 0.01;  $Q=10^{-4}$ for the inset, where the two curves have become indistinguishable.}
\label{fig:ThreeRates}
\endgroup
\bigskip

The three expressions for the coagulation rate constant can now be expressed as follows:
\begin{subequations}
\vspace{-6pt}
\begin{align}
k_S&=4\pi R D,\\
k_A&=k_S\left[1+\frac{2}{\varepsilon}Q\right]^{-1}\label{eq:CoagConstb},\\
k_T&=k_S\left[1+\frac{2-\varepsilon}{\varepsilon}Q\right]^{-1}.\label{eq:CoagConstc}
\end{align}
\end{subequations}
Figure \ref{fig:ThreeRates} shows plots of $k_A/k_S$ and $k_T/k_S$ against $\varepsilon$ for three values of  $Q\equiv\Delta/R$  (0.1 and 0.01 and $10^{-4}$). For $\Delta/R= 10^{-3}$ (not shown in the plot), $k_A/k_S=0.9987$ and $k_T/k_S=0.9993$ at $\varepsilon=1$, and come down to $k_A/k_S=0.9868$ and $k_T/k_S=0.9875$ at $\varepsilon=0.1$. The two curves in the inset are indistinguishable on the scale of the plot, and each stays above 0.99 at $\varepsilon\geq 0.015$, which provides an illustration of the remark made earlier: \textit{ther is no such thing as a grey surface in the continuum world}.

An examination of Fig.\ref{fig:ThreeRates} makes it obvious that if, as is often true for kinetic studies of reactions in solutions, the uncertainty
in the determination of $k_S\equiv 4\pi DR$ exceeds 10\% and if $Q<0.1$, which is also true for ordinary liquids, it is impossible to distinguish between $k_T$ and $k_S$ for values of $\omega\equiv \alpha/(2-\alpha)$ the vicinity of unity. Indeed, more realistic estimates would put $Q\leq 0.05$, which means that, in most cases, one can
only speak of two categories of reactions,  ``diffusion-controlled'' (i.e. , $k_{\rm exp}\approx k_S$) and "activation-controlled" (i.e., $k_{\rm exp}\ll k_S$).

\section{Kolmogrov's partial rediscovery of Burger's approach}

Footnote 1 on p. 202 of Fuchs's 1934 contribution \cite{Fuchs+Y1934+199} refers to an article that is Ref. \cite{Kolmogorov+Lenotovich1933} in our bibliography, and will hereafter be referred to as K\&L1933. What one finds is indeed a highly interesting article on Brownian motion, which falls into three sections: ``In \S1 we explain a method for solving this problem, and in \S\S2, 3 this method is applied to computing the mean Brownian area. \S\S1, 2 of this paper [consisting of formal theory] are written by A. N. Kolmogorov, §3 [where the solution is worked out] by M. A. Leontovich.'' Smoluchowski's expressions for  $n(r,t)$ and the flux are not to be found anywehre, which makes sense, because the authors, who assume that the projection of a Brownian particle diffusing in three-dimensional isotropic space can be represented by a disk, explicitly state: ``In what follows we only consider such projections, so for simplicity we call them particles.'' Nor would one find any mention of the ABC. Fuchs had, it seems, a rather vague understanding of the contents of K\&L1933 at the time of writing his 1934 article ``On the theory of coagulation''. Fortunately, he made amends by presenting, in \textit{The Mechanics of Aerosols} \cite[pp. 188--93]{Fuchs1964Mechanics}, a suitably adapted and detailed account of Kolmogorov's approach, the adaptaion consisting mainly of considering the one-dimensionsal variant of the problem; his symbols $w*$ and $W*$ correspond to Burger's $f$ and $\chi$, his equations (37.20) and (37.21) to Eqs. \ref{eq:B-23} and \ref{eq:B-22}, respectively.

\section{Retrospect and prospects}

Pioneer scientists, like other explorers, often mistake their destination and create, a good deal of confusion, some times very long lasting. Think, for example, of Christopher Columbus, whose error was recognized before long, and partly corrected by the renaming of the New World, but not of its inhabitants. Better navigational tools, in particular accurate chronometry, gradually turned map-making into a science and a fluorishing trade. Since neither science nor trade, nor some other vital interest, is threatened by the confusion between two kinds of Indians, those from India proper and those from the New World, the need for eradicating the residual confusion has not been deemed worth the effort.\par

The scientific explorer Smoluchowski discovered a new territory and traversed it with such a whirlwind speed that he kicked up a mass of slow-settling dust, making it hard for himself and most subsequent visitors to see clearly. Had Burger's work received the attention it deserved, the dust would have been sucked out within a few years, and further progress in the unpolluted atomosphere would probably have been far more rapid. Smoluchowski's nostrum remains as popular as ever, notwithstanding the mass of evidence that has been accumulated against the SBC. Eventually---more than sixty years after the publication of HCB1918---a new framework \textit{was} erected, which elicited the approval of the referees---even those who began by looking askance at the many articles through which these developments were communicated. The salient features of the new outlook, which is consistent with studies of neutron transport and Brownian motin in the presence of absorbing boundaries, will be presented in a sequel to this article.   

Dust particles do settle, and scientists in pursuit of a satisfactory theory, like ants foraging for food, do straighten, more often than not, their tracks, through a cumulatiave effort, by detecting absurdities and discarding illogicalities. A century is evidently not long enough to smoothe out \textit{this} tortuous track.   

I end on a historical note: HCB1918 first came to my notice in 2008, and I prepared the first draft of this article in 2009.

\end{multicols}

\end{document}